\documentclass[12pt]{article}
\usepackage{epsfig}

\def\beq{\begin{equation}}
\def\eeq#1{\label{#1}\end{equation}}
\def\eeqn{\end{equation}}

\def\beqa{\begin{eqnarray}}
\def\eeqa#1{\label{#1}\end{eqnarray}}
\def\eeqan{\end{eqnarray}}

\let\bar=\overbar

\def\D{{\cal D}}

\def\Dslash{\not{\hbox{\kern-4pt $D$}}}
\def\dslash{\not{\hbox{\kern-2pt $\del$}}}

\def\msb{{\bar{\ssstyle M \kern -1pt S}}}

\def\Title#1{\begin{center} {\Large {\bf #1} } \end{center}}

\begin{document}

\Title{The Charm Renaissance: \\ \vspace*{0.2cm} D-physics -- a Selective Review}

\bigskip\bigskip


\begin{raggedright}  

{\it Guy Wilkinson\index{Wilkinson, G.}\\
University of Oxford, \\
Denys Wilkinson Building,\\
Keble Road,\\
Oxford, OX1 3RH, \\
UNITED KINGDOM}
\bigskip\bigskip
\end{raggedright}

\section{Introduction}

In recent years, no doubt because of the success of the $B$-physics
programme at BaBar and Belle, charm has been considered the
poor relation of heavy-quark physics.  This is now changing: while
$B$-physics remains the biggest show in town,  it is no longer the
{\it only} show.  As will be argued, measurements in the $D$-sector
have a vital, although indirect, role to play in pinning down
the value of certain critical parameters of flavour-physics.
Furthermore, the charm system provides a powerful laboratory 
in its own right to search for contributions from non-Standard Model (SM) processes.

Three main reasons can be identified which explain why charm physics is
once more, quite correctly, being perceived as an important and exciting
discipline:
\begin{enumerate}
\item{{\bf Precision CKM Tests}  \\ The success of the $B$-factories and the Tevatron
has meant that CKM unitarity triangle tests are achieving successively higher
levels of precision.  This progress will continue with the LHCb experiment
at CERN.  Although the CKM elements being studied are those accessible in
$B$-decays,  charm turns out to be a vital ingredient in the programme.}
\item{{\bf Charm Mixing and its Legacy} \\ The discovery of $D^0-\bar{D^0}$ oscillations
has been the most exciting event of the past couple of years in high energy physics. 
The higher than expected rate is (arguably) intriguing in its own right,
and points the way forward to searches for CP violation (CPV) in the charm sector.}
\item{{\bf Recent Discoveries in Spectroscopy} \\ The discovery of several missing charmonium states, 
and a number of unexpected and possibly exotic resonances 
(the $X$,$Y$ and $Z$) has rekindled interest in the $c\bar{c}$ system as a 
laboratory for studying QCD.}
\end{enumerate}
In this review we focus on the first two topics.  Useful discussion 
of the third item may be found in~\cite{OLSEN}.

\section{Facilities and Experimental Attributes}
\subsection{Overview}

In reviewing the facilities which have contributed to charm physics studies
in recent years, and are expected to do so in future, three complementary 
strands may be distinguished.   

First are the fixed target experiments, 
most significantly those at Fermilab:  E687, E791 and FOCUS.  Second are
the experiments located on $e^+e^-$ machines.  The majority of results
have come from CLEO, BaBar and Belle,  with the most important source
of $D$-meson production being the $e^+e^-$ contiuum lying under the $\Upsilon(4S)$.
An important special case 
of $e^+e^-$ operation  is the threshold running pursued by CLEO-c at
both the $\psi(3770)$ and at around 4170~MeV, where $D_s$ mesons 
are produced.  The significance of these threshold data is explained in more detail
below.   The BES-III~\cite{BES} experiment is expected to follow the lead of CLEO-c, 
and accumulate perhaps 20 times more data over the coming decade.  In the
more distant future it is hoped that a Super-Flavour Factory~\cite{SUPERB} will be
constructed, which will both increase the charm-from-contiuum sample
of the $B$-factories by 1-to-2 orders of magnitude, and also have the ability
to operate at very high luminosity at threshold.

The third important class of facility in which $D$-meson properties have
been (and will be) studied is that of hadron colliders.   The very
high production cross-section gives rise to enormous statistics.
The Tevatron, and CDF in particular, have published impressive
studies exploiting the very large prompt $D^\ast$ samples that are available.
A recent CDF analysis~\cite{KPICDF} used this source to reconstruct 
around 3 million $D^0 \to K\pi$ events from 1.5~$\rm fb^{-1}$ of data.
This programme will continue at LHCb,
where the possibilities of harnessing secondary charm from $B$-decays
has also been explored and shown to be very promising~\cite{PATRICK}. 
Plans are being
made for an upgraded LHCb experiment~\cite{SUPERLHCB} which will run at around 10 times
the luminosity, and have more efficient triggering, which will therefore
provide still larger datasets.

\subsection{Threshold Running and CLEO-c}

\begin{figure}
\begin{center}
\epsfig{file=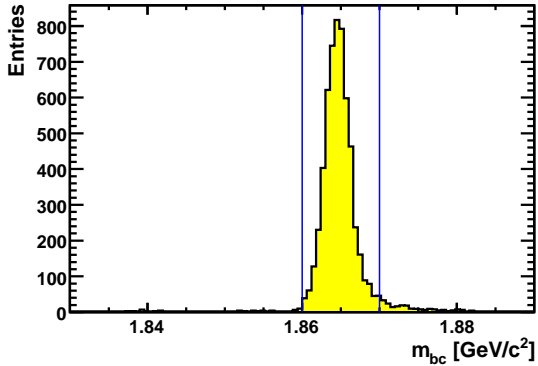,width=0.5\textwidth}
\caption{Beam constrained invariant mass of CLEO-c $D^0 \to K^- \pi^+ \pi^- \pi^+$
events in which the other meson has been reconstructed as $K^+\pi^-$.  (The tail at
high mass arises from signal events in which there has been significant ISR.)}
\label{fig:cleocdtag}
\end{center}
\end{figure}

Threshold running at $e^+e^-$ machines has several attractive characteristics:
\begin{itemize}
\item{The environment is very clean, with no additional fragmentation 
particles produced. This allows for very low backgrounds, particularly 
in the case where both $D$-mesons are reconstructed.  Figure~\ref{fig:cleocdtag}
shows the beam constrained mass plotted in events where one mesons has been
reconstructed in the $K \pi\pi\pi$ and the other as $K \pi$.  The background is at the
level of 1\%.}
\item{Quantum correlation exists between the two $D$ mesons which 
can be exploited.  For example, at the $\psi(3770)$ the quantum numbers
of the resonance mean that if one D meson is reconstructed in a CP-even
state, for example $K^+K^-$, then the CP of the other state is known to 
be CP-odd.  This ability to {\it CP-tag} decays is the most valuable
feature of threshold running, the application
of which is explained in Sec.~\ref{sec:ckm}.}
\item{If all charged particles and photons are reconstructed 
then kinematical constraints allow the the presence of neutral
particles to be inferred, such as $K^0_L$ mesons and neutrinos.
This is useful as it allows CP-tags such as $K^0_L \pi^0$ to be
included in double-tag analyses, and enables the reconstruction 
of  leptonic decays such as $D_s^+ \to l^+ \nu$.}
\end{itemize}
CLEO-c, which completed operation in Spring of this year, accumulated 
818~$\rm pb^{-1}$ of data at the $\psi(3770)$ and 586~$\rm pb^{-1}$ 
at $\sqrt s$~=~4170~MeV.

\subsection{Experimental Attributes}

The desirable attributes needed for a successful $D$-physics experiment
are essentially the same as those required in $B$-physics studies.
These include efficient tracking and, if possible, good calorimetry suitable
for $\gamma$ and $\pi^0$ reconstruction,  hadron identification capabilities
to permit $\pi$-$K$ discrimination, and -- in a hadron collider environment --
a trigger system sensitive to the final states of interest.  The results
discussed in this review come primarily from BaBar, Belle, CLEO-c and CDF -- experiments 
which possess most or all of these characteristics.

\section{$\boldmath D$ Decays and the CKM Unitarity Triangle}
\label{sec:ckm}

\begin{figure}
\begin{center}
\epsfig{file=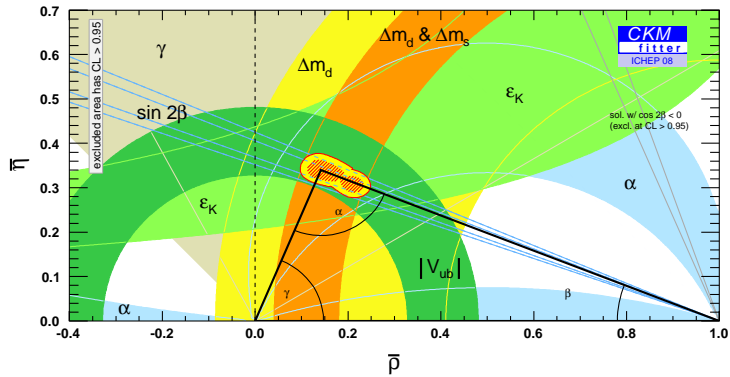,width=0.7\textwidth}
\caption{Constraints in the $\bar{\rho},\bar{\eta}$ plane as of Summer 2008~\cite{CKMFIT}.}
\label{fig:ckm}
\end{center}
\end{figure}

Our understanding of CP-violating phenomena, as expressed in the
context of the Standard Model, is most usefully represented by constraints in
the $(\bar{\rho},\bar{\eta})$ plane, where at order $\lambda^2$, these symbols 
represent two of the
parameters of the CKM-matrix in the Wolfenstein 
parameterisation~\cite{WOLF} multiplied by the factor $(1-\lambda^2)$,
$\lambda$ being the sine of the Cabibbo angle.  All these constraints
(with the exception of $\epsilon_K$, the CP-violating parameter
obtained from kaon decays) come from measurements of $B$-meson 
properties, which are expected to map out a triangle with
vertices $[(0,0), (1,0), (\bar{\rho},\bar{\eta})]$  The present experimental status is summarised in
Fig.~\ref{fig:ckm}.  All measurements are broadly consistent
with each other, indicating the validity of the CKM paradigm. 
Nonetheless, new physics contributions are not excluded
and may become apparent when the experimental precision
improves still further -- this is one of the principal goals of the flavour
physics programme.   In surveying where improvement is necessary
it is natural to focus on both the angle $\gamma$, indicated
in Fig.~\ref{fig:ckm}, and the so-called `mixing' side opposite
to this angle.   The geometry of the triangle means that these two quantities are closely 
linked, for it is the length of the side which largely determines the
expected value of $\gamma$.  
Two possible central values are predicted for the value of $\gamma$ at the one sigma level:  
 $(55.4^{+2.5}_{-2.2})^\circ$ or $(67.4^{+3.3}_{-5.6})^\circ$~\cite{CKMFIT}.
Comparison of the measured and expected values of $\gamma$ is
a powerful way to search for new physics.
For both quantities it will be seen that, despite the
fact that both of these features are measured in $B$-decays,
crucial input is provided by analyses of $D$-decay properties. 
 
\subsection{The CKM Angle $\gamma$ and $\boldmath D$-decays}

The most powerful manner in which to measure the angle $\gamma$
is with $B^\pm \to D K^\pm$ decays.  Here two tree diagrams contribute,
one of which involves a $D^0$
meson and the other a $\bar{D^0}$ meson.   If a final state is chosen which
is common to both $D^0$ and $\bar{D^0}$ then interference 
occurs that includes terms dependent on the phase difference
between the diagrams,
which is $\delta_B - \gamma$, where $\delta_B$ is a CP-conserving
strong phase.  Comparing suitable observables between 
$B^-$ and $B^+$ decays allows $\gamma$ and $\delta_B$ to be determined, along
with $r_B$, a parameter which represents the relative strength of the two diagrams 
($\approx 0.1$).   Categories of $D$-decays which have been proposed for
these measurements include CP-eigenstates~\cite{GLW}, for example $K^+K^-$ or $K^0_S \pi^0$, 
Cabibbo favoured and doubly-Cabibbo
suppressed decays (the so-called `ADS' approach~\cite{ADS}), for example $K^-\pi^+$ or $K^-\pi^+\pi^-\pi^+$, 
and self-conjugate multibody states, such as $K^0_S \pi^+\pi^-$~\cite{GGSZ}.  It is
this latter method which has yielded the best constraints on $\gamma$ with
the statistics presently available at the $B$-factories~\cite{BABARKSPP,BELLEKSPP}.

\begin{figure}
\begin{center}
\epsfig{file=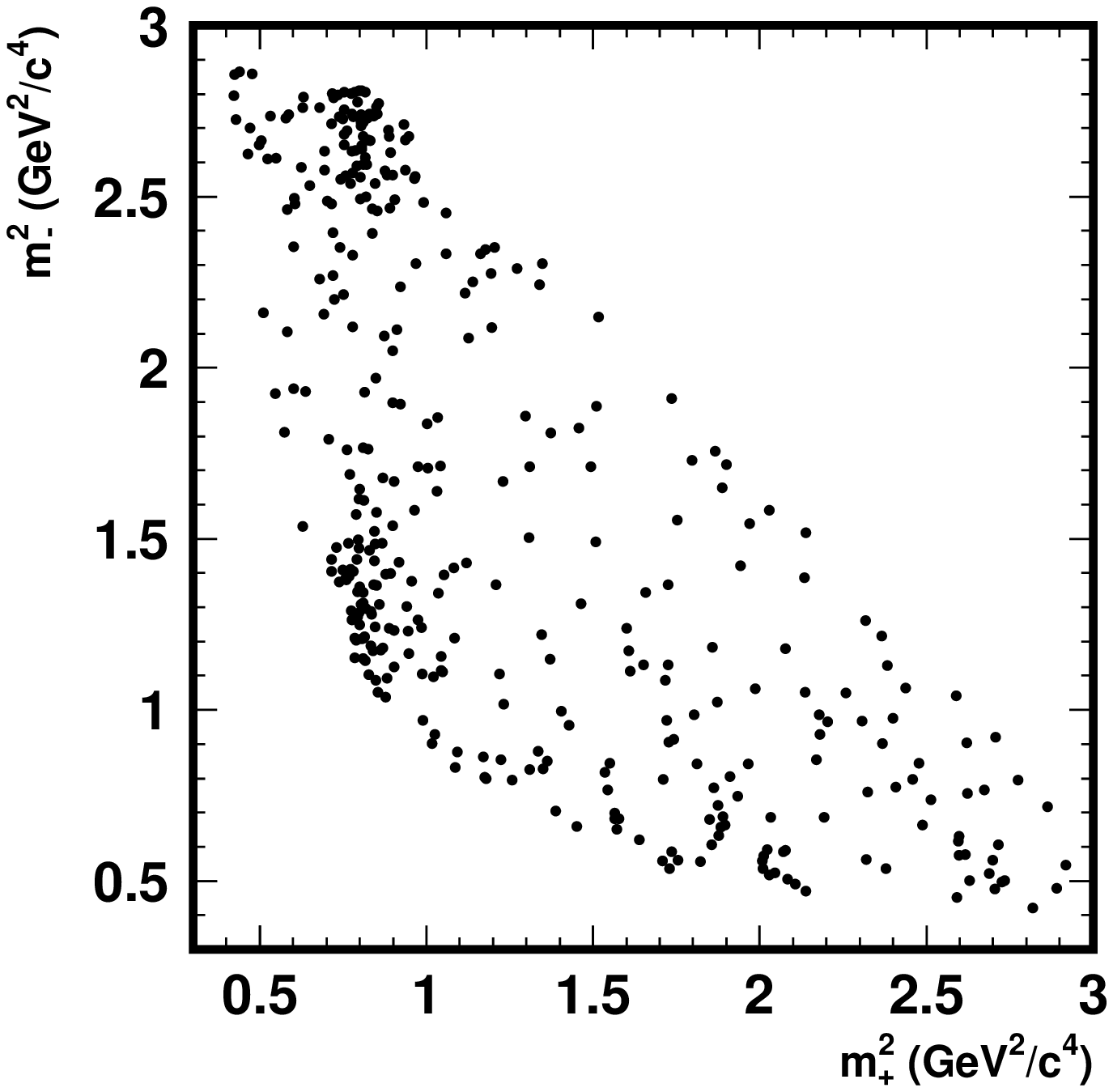,width=0.45\textwidth}
\epsfig{file=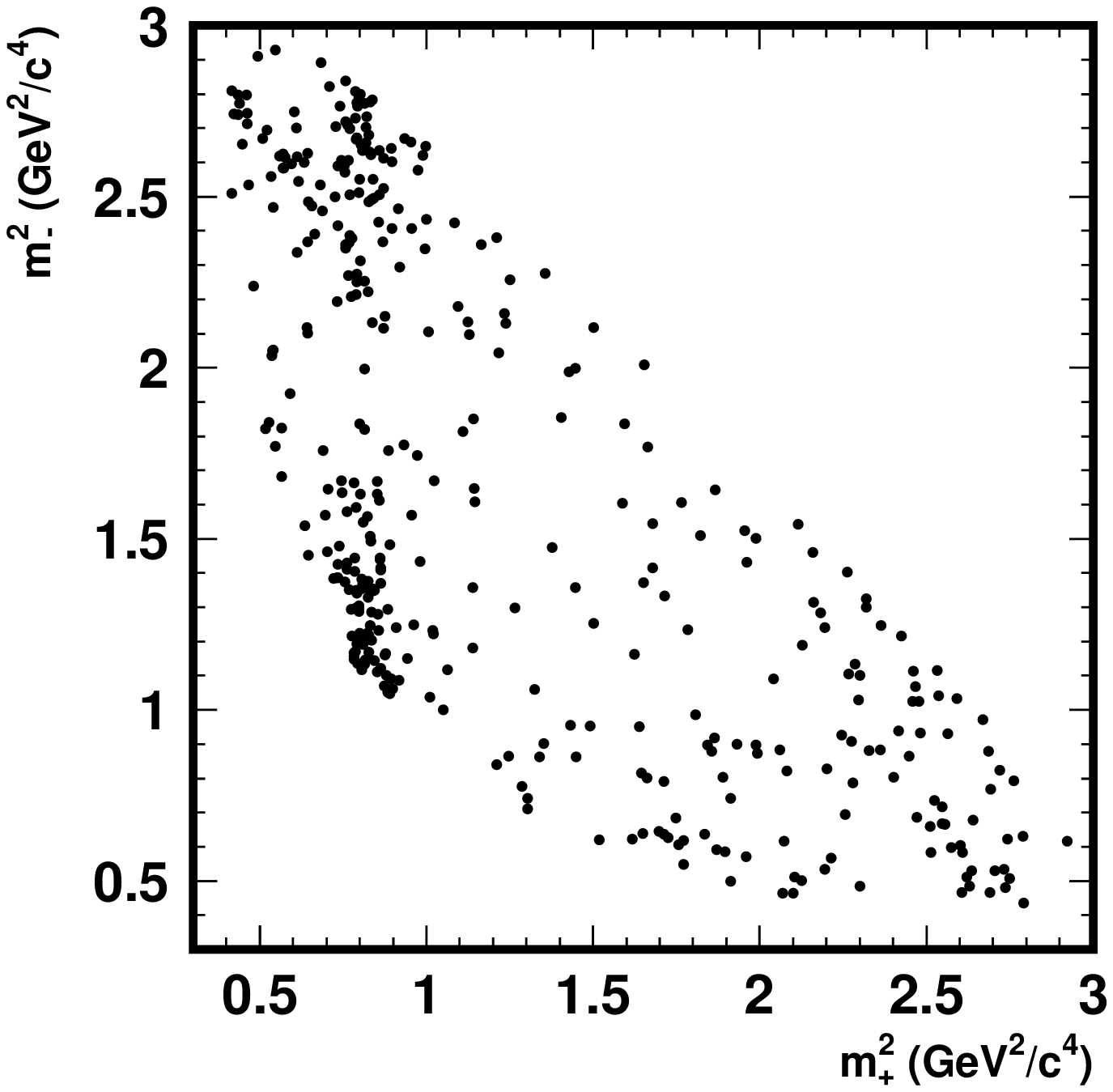,width=0.45\textwidth}
\caption{Dalitz plots of $D^0 \to K^0_S \pi^+\pi^-$ decays arising 
from the process $B^\pm \to D K^\pm$~\cite{BELLEKSPP}.  Left: $B^-$; right: $B^+$.
The horizontal axis is the invariant mass squared for the $K^0_S \pi^+$ pair,
and that of the vertical axis the same for the $K^0_S \pi^-$.}
\label{fig:belledalitz}
\end{center}
\end{figure}

In using $D^0 \to K^0_S \pi^+\pi^-$ decays in a $B^\pm \to D K^\pm$ analysis
the CP-sensitive observable is the Dalitz plot of the $D$-decay.
A non-zero value of $\gamma$ will give rise to differences in the distributions for
$B^+$ and $B^-$ decays.  Examples plots from a recent BELLE analysis~\cite{BELLEKSPP} are
shown in Fig.~\ref{fig:belledalitz}.  If the composition of the 
intermediate resonances involved in the $D$-decay is understood, then
a comparison of the two plots allows $\gamma$ to be extracted through
an unbinned likelihood fit.  In this manner
charm physics provides a critical input to the $\gamma$ measurement.

\begin{figure}
\begin{center}
\epsfig{file=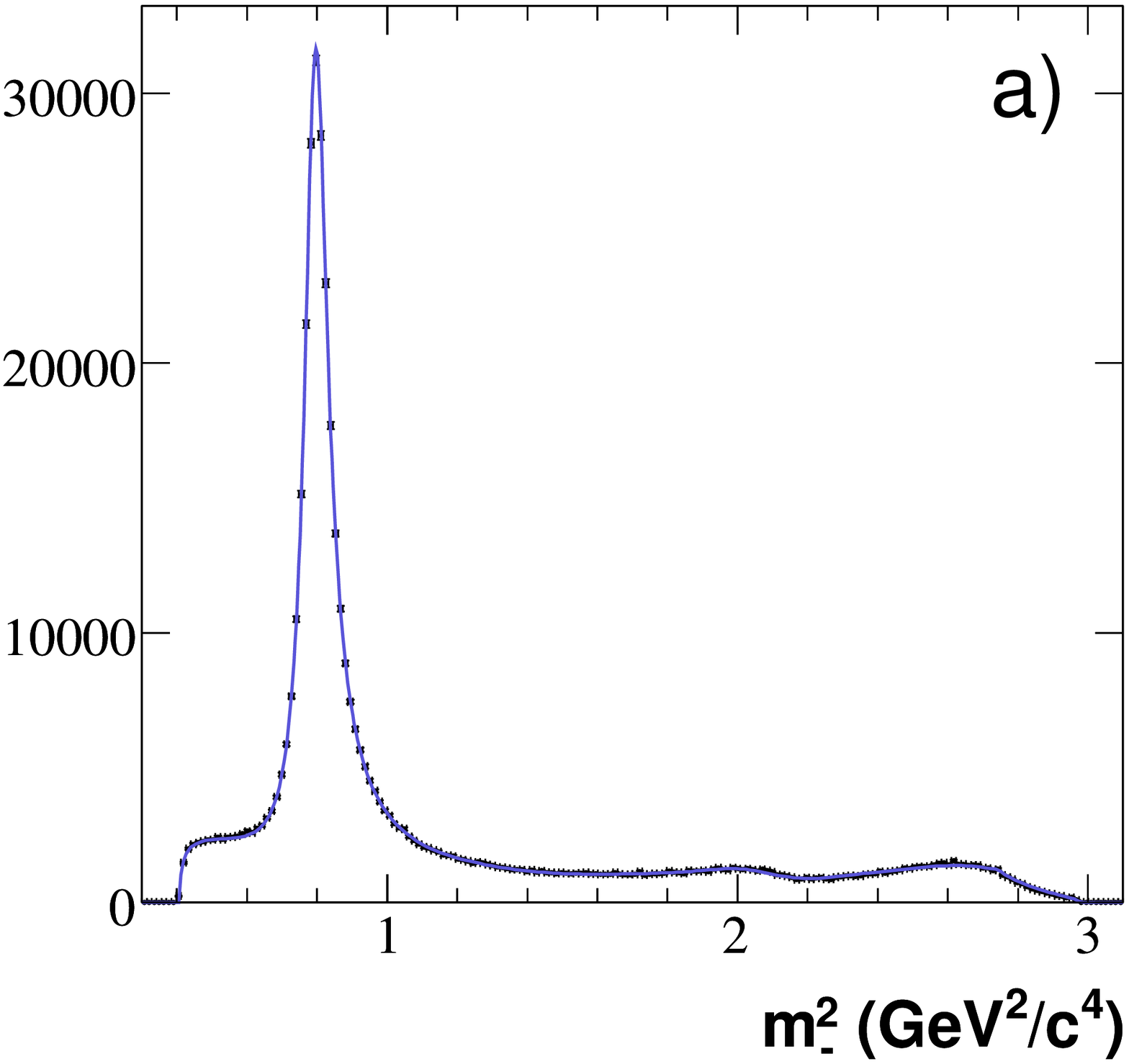,width=0.32\textwidth}
\epsfig{file=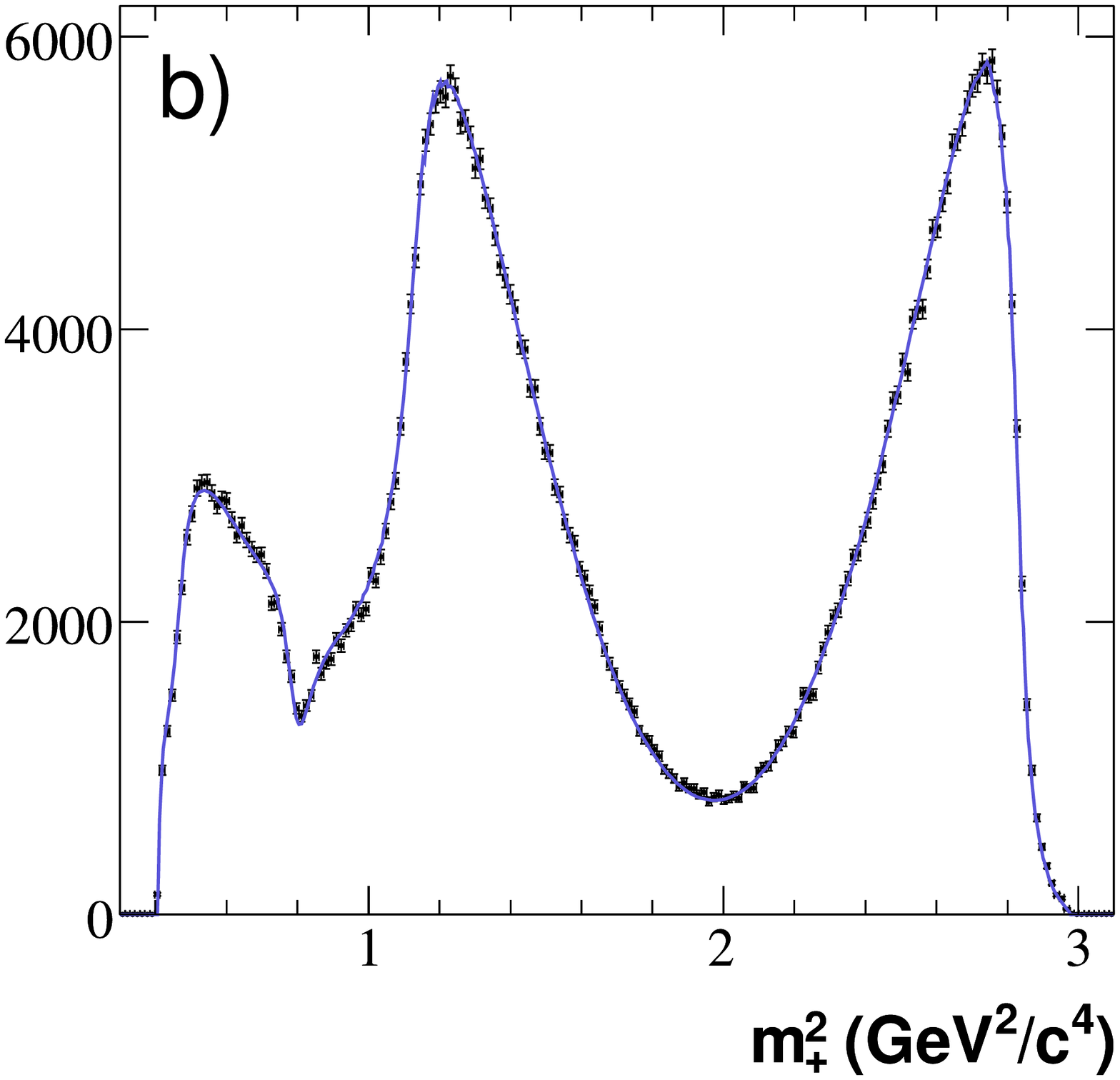,width=0.315\textwidth}
\epsfig{file=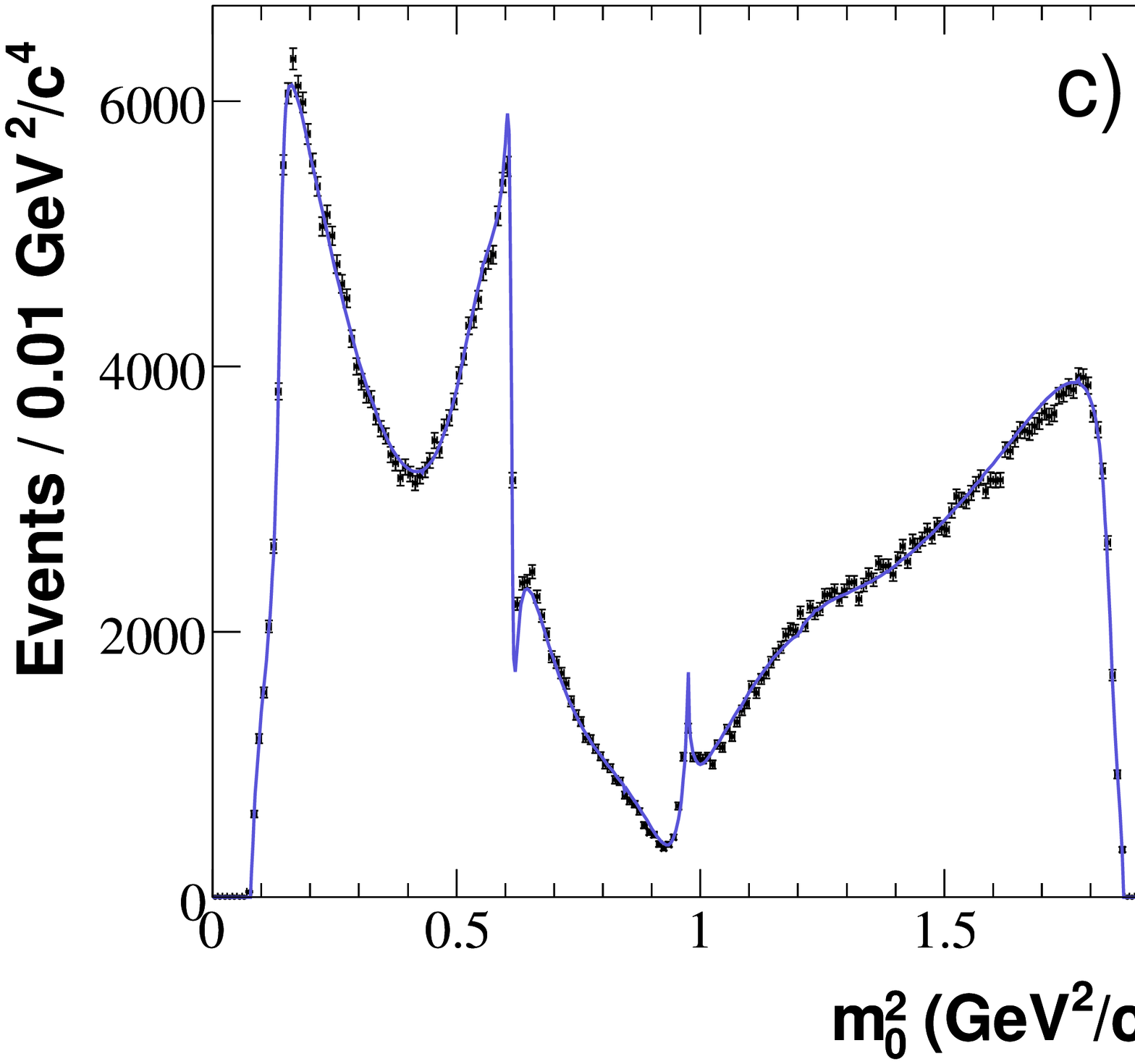,width=0.32\textwidth}
\caption{BELLE projections of the Dalitz variables for the decay $D^0 \to K^0_S \pi^+\pi^-$
with fit result superimposed~\cite{BABARKSPP}. Invariant mass squared for the: $K^0_S \pi^-$ (a),
the $K^0_S \pi^+$ (b) and the $\pi^+ \pi^-$ (c).}
\label{fig:ksppproj}
\end{center}
\end{figure}

The $B$-factory experiments have devoted a great deal of effort to modelling the
$D^0 \to K^0_S\pi^+\pi^-$ decay for the purposes of the $\gamma$ measurement.
A recent BaBar study~\cite{BABARKSPP} has used a sample of 487k flavour tagged
$D^{*\pm} \to D \pi^\pm$ events to which an isobar model involving
ten resonances is fitted.  The $\pi\pi$ and $K^0_S\pi$ S-wave contributions are described
with a K-matrix~\cite{KMATRIX} and LASS~\cite{LASS} parametrisation respectively.
Projections of the Dalitz plot, with the model fit superimposed, are shown in
Fig.~\ref{fig:ksppproj}.  The $\chi^2$ of the fit is 1.11 for 19274 degrees of freedom.
In the $\gamma$ fit a systematic uncertainty is incurred arising from 
how well this model represents reality.  This error
is assigned to be $7^\circ$, which is small compared with the statistical
uncertainty, but will become limiting with the higher statistics
$B$-samples expected at LHCb.   

\begin{figure}
\begin{center}
\epsfig{file=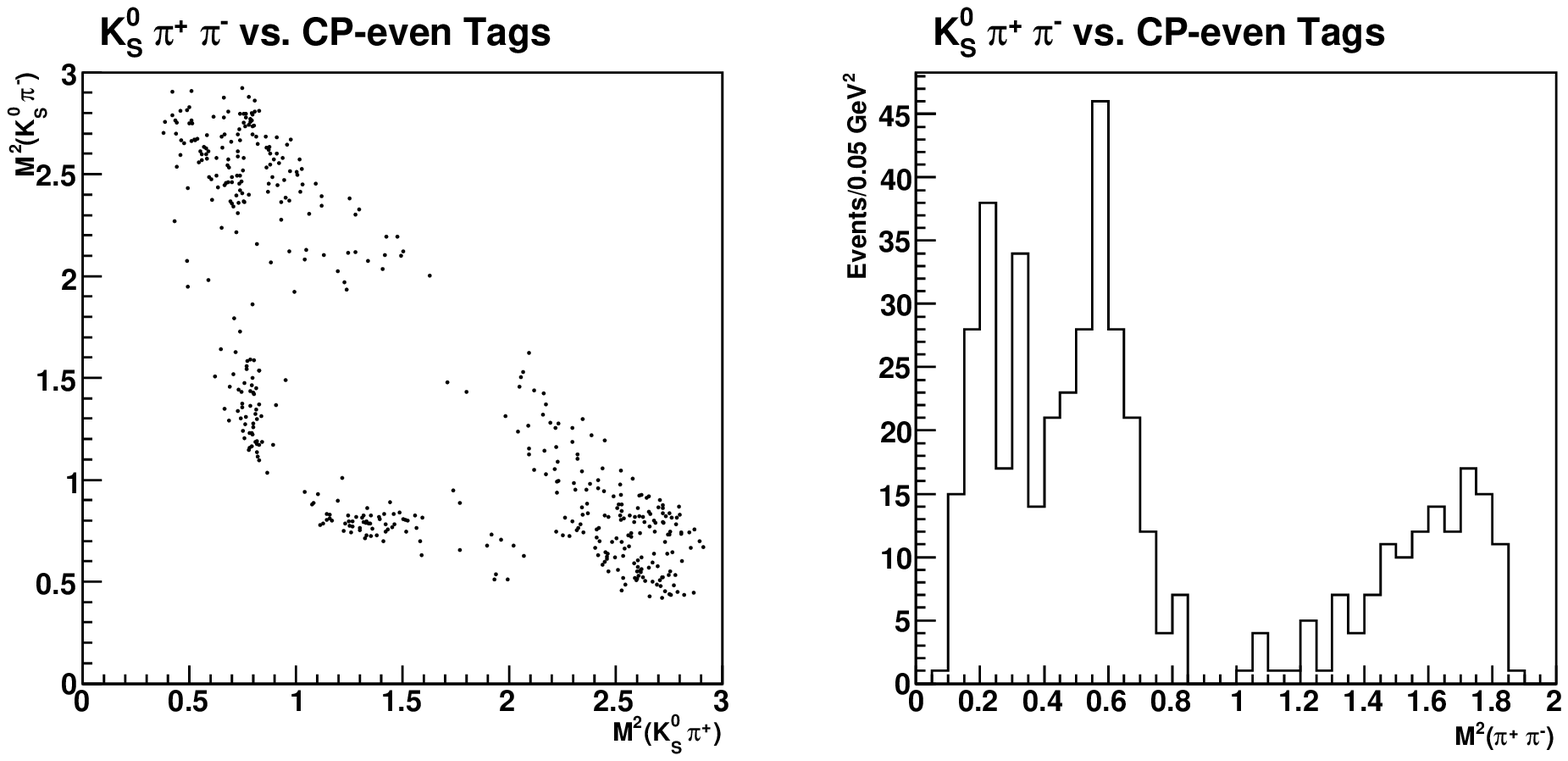,width=0.75\textwidth}
\epsfig{file=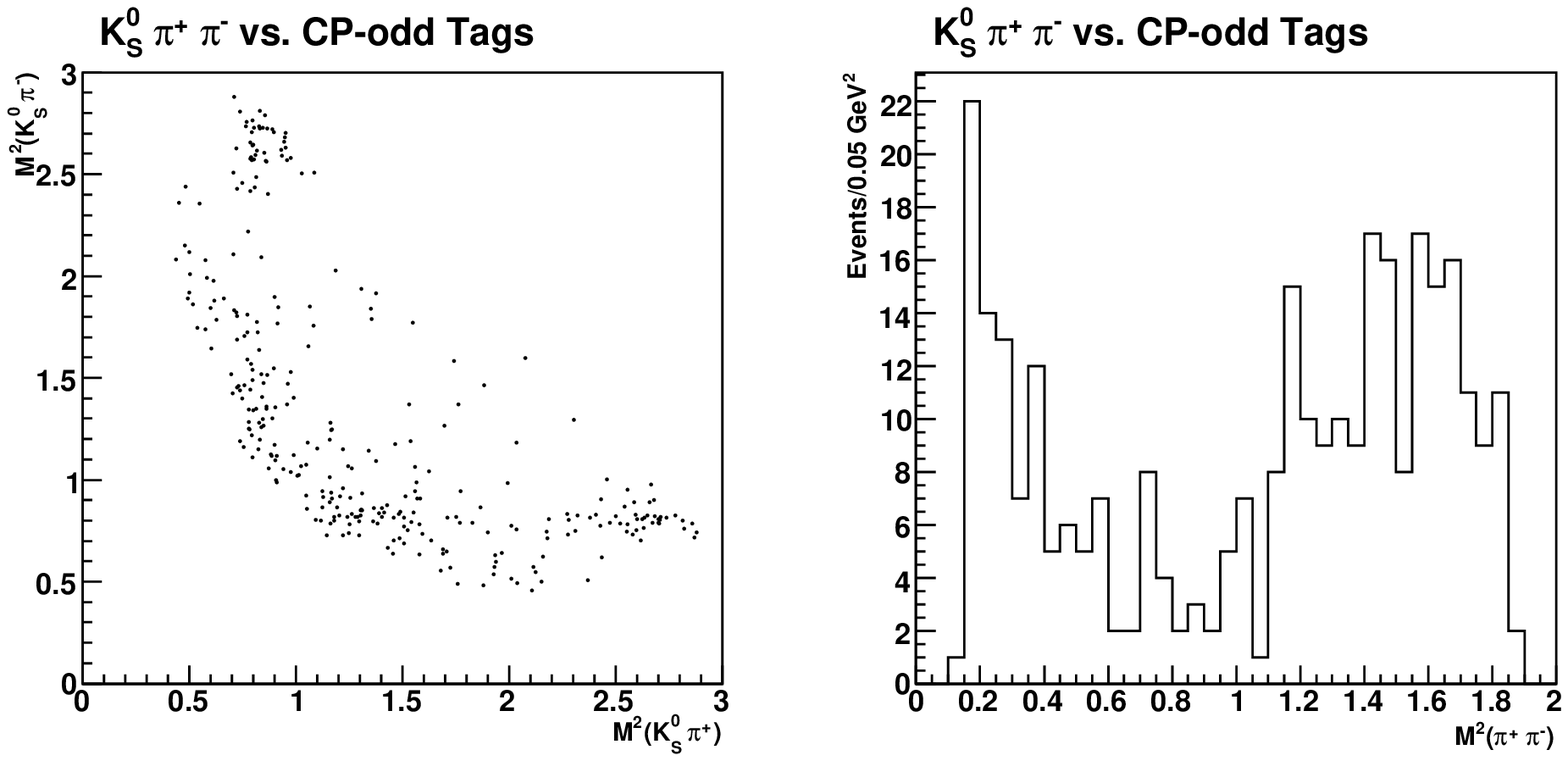,width=0.75\textwidth}
\caption{CP-tagged $D^0 \to K^0_S \pi^+\pi^-$ Dalitz plots and $\pi^+\pi^-$ 
projections from CLEO-c.}
\label{fig:cptags}
\end{center}
\end{figure}

An alternative approach is to make a binned fit
in which the model predictions are replaced by quantities which are directly
measured in double-tagged quantum-correlated D-decays at the $\psi(3770)$~\cite{BONDAR}.
If one D-meson is reconstructed in a CP-eigenstate then the other meson will
be in the opposite eigenstate, that is a known superposition of $D^0$ and $\bar{D^0}$.
So if this meson is reconstructed as $K^0_S\pi^+\pi^-$ then there will be contributions from
$D^0$, $\bar{D^0}$ and an interference term involving the strong phase difference
between the two decay paths.  It is this information which is invaluable in the
$\gamma$ measurement and is inaccessible through direct means in flavour-tagged D-decays.
Similarly useful input comes from events containing two $K^0_S\pi^+\pi^-$ decays.
Quantities required for the $\gamma$ extraction (the so-called $c_i$ and $s_i$ coefficients -- 
see~\cite{BONDAR}) are measured which are directly related
to the relative population of the chosen bins for different combinations of tags.
Figure~\ref{fig:cptags} shows $K^0_S\pi^+\pi^-$ Dalitz plots and the corresponding
projections made with CLEO-c data for CP-even and CP-odd tags.  The difference in structure 
is apparent, for example the absence of the $K^0_S\rho^0$ peak in the events containing 
a CP-odd tag.   Preliminary results exist from CLEO-c~\cite{CKMJONAS};  the finite $\psi(3770)$
sample size will induce a residual error of $1-2^\circ$ on $\gamma$.
Although the binned treatment leads to some degradation in $B$-statistical precision, it 
is still expected at LHCb that this model independent approach will outperform the 
model dependent fit after one-year (2~${\rm fb^{-1}}$) of data-taking~\cite{MYCKM}, 
with an error dominated by measurement uncertainties alone. 

There are other equally important ways in which quantum-correlated $D$-decays can
be harnessed for the $B^\pm \to DK^\pm$ $\gamma$ analysis, and which are being
explored on CLEO-c.  These include determinations of the strong phase difference
in $D^0 \to K-\pi^+$ decays~\cite{CLEODKP}, and the {\it coherence factor}~\cite{COHERENCE} 
and average strong phase difference 
in $D^0 \to K^-\pi^+\pi^-\pi^+$ and $D^0 \to K^-\pi^+\pi^0$ decays~\cite{CKMJIM},
all of which are important for the ADS family of $\gamma$ measurements.

An average of existing measurements with $B$-data yields $\gamma =(67^{+32}_{-25})^\circ$~\cite{CKMFIT}.
LHCb has the potential of reducing this uncertainty to $2-3^\circ$~\cite{LHCBGAM},
using a combination of methods, the majority of which will rely critically on
the knowledge of the $D$-meson decay structure.

\subsection{The `Mixing Side' and D-Meson Tests of Lattice QCD}
\label{sec:lattice}
 
The ratio $|V_{td}/V_{ts}|$ can be used to fix the  unitarity triangle side opposite to the angle $\gamma$.
This ratio is  determined from the ratio of oscillation frequencies $\Delta m_{B_s}$ to $\Delta m_{B_d}$ in the 
$B^0$ and $B_s$ systems:
\begin{equation} 
|V_{td}/V_{ts}| = (f_{B_s}\surd B_{B_s})/(f_{B_d}\surd B_{B_d})\surd(\Delta m_{B_d} \, m_{B_d} / \Delta m_{B_s} 
\, m_{B_s}).
\end{equation}
Also involved are the meson masses, $m_{B_{d(s)}}$, the meson decay constants $f_{B_{d(s)}}$ and 
the bag factors $B_{B_{d(s)}}$.   The ratio $(f_{B_s}\surd B_{B_s})/(f_{B_d}\surd B_{B_d})$ is 
calculated in lattice QCD~\cite{MIXLATTIC} to be $1.23 \pm 0.06$. It is the uncertainty in this calculation
which dominates the knowledge of the length of the side, and hence the prediction for the expected 
value of $\gamma$.   For this reason, it is highly desirable to make experimental validations of the
lattice QCD calculations.  This cannot readily be done in the B-system,  but it is possible to make
measurements of $f_D$ and $f_{D_s}$, the form factors for $D^+$ and $D_s$ decays.  Comparison 
between these results and the lattice calculations then provide a critical test of the lattice approach.

The $D^+$ and $D_s^+$ form factors may be measured from leptonic meson decays.   The partial width
for a $D^+$ decaying to $l^+\nu$ is given by
\begin{equation}
\Gamma(D^+\to l^+\nu) = \frac{1}{8\pi} G_F^2 f_D^2 m_l^2 m_D \left( 1 - \frac{{m_l}^2}{{m_D}^2} \right)^2 
|V_{cd}|^2
\end{equation}
and similarly for $D_s^+$, but here involving the parameters 
$f_{D_s}$, $m_{D_s}$ and $|V_{cs}|$.  If the values for the 
magnitudes of the CKM angles are taken from elsewhere, the form factors may be extracted.
Measurements of $f_{D_s}$  come from CLEO-c~\cite{CLEOCFDS1,CLEOCFDS2}, BELLE~\cite{BELLEFDS} and
BaBar~\cite{BABARFDS}, and a recent new determination  of $f_D$ has been made by CLEO-c~\cite{CLEOCFD}.
These are to be compared with the most precise available lattice calculation from~\cite{FOLLANA}.

\begin{figure}
\begin{center}
\epsfig{file=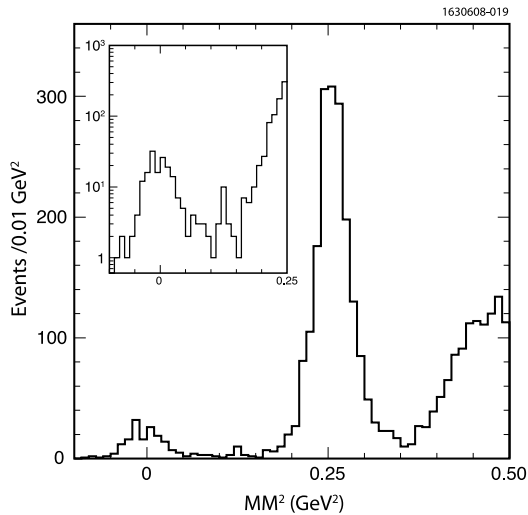,width=0.45\textwidth}
\caption{Missing mass squared in the CLEO-c $D^+ \to \mu \nu$ analysis~\cite{CLEOCFD}.}
\label{fig:cleofd}
\end{center}
\end{figure}

The CLEO-c measurements are based on full-reconstruction techniques which exploit
the cleanliness of the threshold environment.  In the $f_D$ analysis, for example,
events are considered from the $\psi(3770)$ running where one charged $D$ meson is found 
together with a single other charged track, which is minimum ionising.  The missing 
mass for these events is shown in Fig.~\ref{fig:cleofd}.  A peak is seen
at zero-missing mass, consistent with $D^+ \to \mu^+\nu$ events and clearly
separated from the background process $D^+ \to K^0 \pi^+$.  A small contribution
is seen between these two peaks coming from $D^+ \to \tau^+(\pi^+\bar{\nu})\nu$.
Using the population of this peak to measure the branching ratio yields a result
$BR(D^+\to\mu^+\nu)= (3.82 \pm 0.32 \pm 0.09) \times 10^{-4}$ and a form-factor value
which is $f_D = (205.8 \pm 8.5 \pm 2.5)$~MeV (this result is with the ratio
of $\tau^+\nu$ and $\mu^+ \nu$ decays fixed to the Standard Model expectation)~\cite{CLEOCFD}.  
Similar methods are used for the $f_{D_s}$ measurement at CLEO-c based on the 4170~MeV dataset,
using both muon and tau decays~\cite{CLEOCFDS1,CLEOCFDS2}.  
In~\cite{CLEOCFDS1} $D_s^+ \to \mu+ \nu$ and $\tau^+(\pi^+\nu)\nu$ decays
from 314~$\rm pb^{-1}$ of data are used to determine $B(D_s^+ \to \mu^+\nu) = (8.00 \pm 1.3 \pm 0.4) \times 
10^{-3}$ and $f_{D_s} = (274 \pm 13 \pm 7 )$~MeV.
     
At the B-factories the procedure is first to infer the presence of a $D_s^+$ from
the recoiling mass seen against the system of a reconstructed $D$ and fragmentation particles,
and then to look for a muon and compute the invariant mass of what remains. With 548~${\rm fb^{-1}}$ of
data Belle  have measured $B(D_s^+ \to \mu^+\nu) = (6.44 \pm 0.76 \pm 0.57) \times 10^{-3}$, implying 
a form factor result of $f_{D_s} = (275 \pm 16 \pm 12 )$~MeV.

The CLEO result for $f_D$ is in agreement with the lattice calculation of $f_D = 207 \pm 4$~MeV~\cite{FOLLANA}.
The experimental results for $f_{D_s}$ are consistent with each other and give a result of 
$f_{D_s} = 270 \pm 8$~MeV, which is three sigma above the lattice value of $f_{D_s} = 241 \pm 3$~MeV.
This is an intriguing situation, which could possibly hint at problems in the lattice approach,
or even new physics contributions to the $D_s$ decay~\cite{KRONFELD}.  It is therefore imperative
to improve still further the experimental precision.  New results for $f_{D_s}$ are expected soon
from CLEO-c with the full 4170~MeV dataset.  Updates are also possible from BaBar and Belle.
BES-III data will allow for improved measurements of both $f_{D_s}$ and $f_D$.

It is also possible to measure semi-leptonic form factors in $D$ decays~\cite{SEMILEP}, which
then allow for another test of lattice QCD predictions.

\section{Searches for New Physics in Charm Mixing \\ and Decays}

The recent discovery of mixing in the $D^0$ system, after 30 years of experimental 
effort, is a significant milestone in flavour physics.  Here a very brief summary
is given; a full review can be found elsewhere at this conference~\cite{MARKS}.
Attention is now turning to the search for CP violation in the charm sector,
which is an outstanding method to probe for evidence of contributions from
non-SM processes.  Rare charm hadron decays, although not in general
allowing for the cleanliness of interpretation that is familar in B-physics,
constitute another area in which beyond-the-SM physics may manifest itself.  It must be
emphasised that the processes discussed here, suppressed or forbidden in
the SM, offer a complementary route to the search for new physics
to those pursued elsewhere in flavour physics.  In contrast to the kaon and B-meson
sectors, rare charm transitions are unique in receiving contributions from loop
diagrams involving virtual down-type quarks.  

\subsection{Charm Mixing and CP Violation}

$D^0-\bar{D^0}$ transitions are governed by the two parameters
\begin{equation}
x = \frac{\Delta M}{\Gamma} \hspace{0.5cm} {\rm and} \hspace{0.5cm} y = \frac{\Delta \Gamma}{2\Gamma},
\end{equation}
where $\Delta M$ and $\Delta \Gamma$ are the mass and width differences respectively between the
two mass eigenstates, and $\Gamma$ the mean width of these eigenstates.  Mixing can
be mediated by short or long-distance processes.    GIM suppression 
and the values of the relevant CKM elements make the short-distance contributions tiny
in the Standard Model.  Box diagrams alone are expected to lead to values of $x \sim 10^{-5}$ and $\sim 
10^{-7}$~\cite{FALK}.  Long distance effects can however be sizable, particularly for $y$ where they
are expected to be dominant, and able to enhance these values by many orders of magnitude.  The only
clear signature of new physics therefore would be the observation  $x>>y$.

Since 2006 the B-factories have produced results of high precision in a range
of complementary strategies sensitive to $D^0-\bar{D^0}$ mixing.  These have
included searches for mixing in the interference between Cabibbo-favoured  and doubly-Cabibbo
suppressed decays~\cite{KPIBFACT} (with an impressive result in the same analysis also
emerging from the Tevatron~\cite{KPICDF}), measurements of the lifetime in decays to 
CP-eigenstates~\cite{YCPBFACT},
and mixing-sensitive amplitude analyses of $D^0 \to K^0_S \pi^+\pi^-$ decays.  Although no single
measurement has yet yielded a five-sigma discovery in isolation,  the cumulative evidence 
is beyond doubt.    A global fit of all results (allowing for CPV) 
excludes the no-mixing hypothesis at 9.8~$\sigma$ and yields parameter values of 
$x=1.00^{+0.24}_{-0.26}$ and $y=0.76^{+0.17}_{-0.18}$~\cite{HFAGCHARM}.
Although these values are larger than many commentators expected~\cite{MIXPRED},
they remain in accordance with SM expectations.  Nevertheless, the absence
of a clear indicator of beyond-the-SM effects can be used to set constraints on
a host of new physics models~\cite{GOLOWICH}.  Furthermore, the larger-than-anticipated
value of the oscillation parameters is encouraging in the 
search for mixing-related CPV.

\begin{figure}
\begin{center}
\epsfig{file=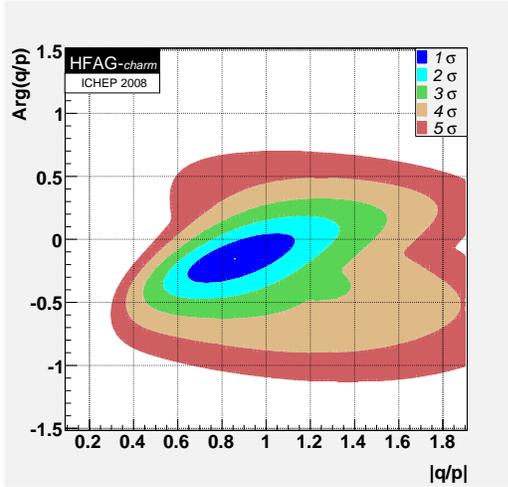,width=0.45\textwidth}
\caption{World average results for $|q/p|$ and $\phi$~\cite{HFAGCHARM}.}
\label{fig:qpphi}
\end{center}
\end{figure}

An unambiguous signature of the existence of new physics processes would be the
discovery of CPV, either in the mixing, or in the interference between
mixing and decay.  The former effect is characterised by a non-zero 
value of $\epsilon = |q/p|-1$, and the latter by a finite value of 
$\phi=\arg({q/p})$, where $q$ and $p$ are the coefficients
which relate the mass ($|D_{1,2}>$) and flavour ($|D^0>$, $|\bar{D^0}>)$ eigenstates:
$|D_{1,2}>\,=\, p|D^0>\,\pm \,q|\bar{D^0}>$.  In the SM both $\epsilon$ and $\phi$ 
are negligibly small~\cite{BIGI}.  Sensitivity to these parameters is achieved
by merely generalising the fits of the mixing analyses to allow for CPV contributions.
A global average of the present results, displayed in Fig.~\ref{fig:qpphi},
shows no indication of CPV, with one sigma uncertainties on $\phi$ and
$\epsilon$ of $0.13$ and $0.16$ respectively~\cite{HFAGCHARM}.  This precision is already
an impressive achievement given the short history of $D^0$-mixing studies.
Improvements at LHCb and future facilities are eagerly anticipated.

One may look for the effects of direct CPV in a final state $f$ by 
searching for non-zero values of the asymmetry $A^f_{CP}$:
\begin{equation}
A^f_{CP} = \frac{\Gamma(D\to f) \, - \, \Gamma(\bar{D} \to \bar f)}
{\Gamma(D\to f) \, + \, \Gamma(\bar{D} \to \bar f)}.
\label{eq:afcp}
\end{equation}
In the case that the decaying meson is a $D^0$ then 
the result includes possible contributions
from mixing and mixing-induced CPV, as well as the direct component.
Singly-Cabibbo suppressed decays are the most interesting place to 
look for direct CPV, as gluonic penguins lead to the possibility of
significant effects (up to $10^{-2}$) from many new 
physics models~\cite{NIRETAL}, and indeed to non-negligible contributions
from the SM itself~\cite{BIGI}. 

\begin{figure}
\begin{center}
\epsfig{file=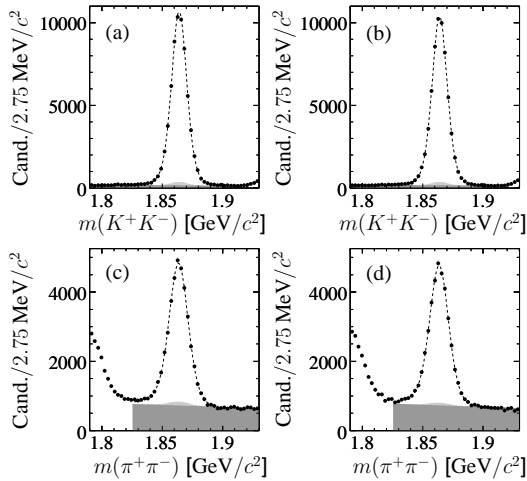,width=0.45\textwidth}
\caption{
Mass-peaks from BaBar $A^f_{CP}$ analysis~\cite{BABARCPVKK}.  (a) and (b) show the
$D^0$ and $\bar{D^0}$ peaks from the $K^+K^-$ analysis, and
(c) and (d) the corresponding peaks from the $\pi^+\pi^-$ analysis.
Solid grey indicates non-peaking background; light grey peaking background.}
\label{fig:acpkkpp}
\end{center}
\end{figure}

Recently there have been results~\cite{BABARCPVKK,BELLECPVKK,BABAR3BOD,BELLE3BOD,CLEODKKP} 
which are achieving sub-percent precision on measurements of $A^f_{CP}$ for singly-Cabibbo
suppressed decays.  These results are summarised in Table~\ref{tab:scsacp}. 
Although all results are consistent with zero CPV, it is clear that experiments are
now entering a very interesting regime.
Figure~\ref{fig:acpkkpp} 
shows the $D^0$ and $\bar{D^0}$ mass peaks from the BaBar $K^+K^-$ and $\pi^+\pi^-$ analyses,
which have around 130k and 64k signal events respectively. 
What is 
impressive about these analyses is the manner in which the systematic uncertainties
have been controlled.  In the $B$-factory $K^+K^-$ and $\pi^+\pi^-$ 
analyses~\cite{BABARCPVKK,BELLECPVKK}, for example, $D^0 \to K^-\pi^+$ events have been 
used to calibrate out detector asymmetries associated with the `slow pion'
in the $D^\ast$ reconstruction, and care has also been taken to remove the effect
of the forward-backward asymmetry coming from the $\gamma-Z$ interference in the
$e^+e^-$ annihilation.

\begin{table}
\begin{center}
\begin{tabular}{l|rrr}  
Mode & \multicolumn{1}{c}{BaBar} & \multicolumn{1}{c}{BELLE} & \multicolumn{1}{c}{CLEO} \\ \hline
$K^+K^-$     & $0.00 \pm 0.34 \pm 0.13$~\cite{BABARCPVKK} &
               $-0.43 \pm 0.30 \pm 0.11$~\cite{BELLECPVKK} & \\ 
$\pi^+\pi^-$ & $-0.24 \pm 0.52 \pm 0.22$~\cite{BABARCPVKK} &
               $0.43 \pm 0.52 \pm 0.12$~\cite{BELLECPVKK} & \\
$K^+K^-\pi^0$& $1.00 \pm 1.67 \pm 0.25$~\cite{BABAR3BOD} & &
\\
$\pi^+\pi^-\pi^0$ & $0.62 \pm 1.24 \pm 0.28$~\cite{BABAR3BOD} &
                   $0.43 \pm 0.41 \pm 1.23$~\cite{BELLE3BOD} &
				   \\
$K^+K^-\pi^+$ &  &  & $-0.03 \pm 0.84 \pm 0.29$~\cite{CLEODKKP} \\ \hline
\end{tabular}
\caption{Recent results in percent for $A^f_{CP}$ in singly-Cabibbo suppressed $D$ decays.
The uncertainties are statistical and systematic respectively.}
\label{tab:scsacp}
\end{center}
\end{table}
	
In the case that the decays under study involve more than two particles,
then the analysis may be extended to consider
final state distributions, which can in principle be more sensitive
to CPV than the overall rates.   BaBar have done this in a comprehensive manner in their $K^+K^-\pi^0$
and $\pi^+\pi^-\pi^0$ study~\cite{BABAR3BOD}, using a variety of techniques
to search for differences between $D^0$ and $\bar{D^0}$ for the Dalitz plots
of the final state particles.   In four body decays one 
can pursue analogous methods or study triple-product correlations~\cite{BIGI}.  A pilot
analysis using the latter approach has been performed by FOCUS for the decay $K^+K^-\pi^+\pi^-$~\cite{FOCUS}.
No signal is yet seen of CPV.

\subsection{Rare Charm Decays}

The most interesting rare charm decay is the process $D^0\to \mu^+\mu^-$.  
In the SM this is extremely suppressed ($\sim 10^{-13}$) but it can be
dramatically enhanced in R-parity violating SUSY, which
allows for branching ratios up to the level of $10^{-6}$~\cite{RARECHARM1}.
The best existing limit comes from CDF which excludes this decay down
to a rate of $4.3 \times 10^{-7}$ at the 90\% C.L.~\cite{CDFD0MUMU} using
360~$\rm pb^{-1}$ of data.

As it the case for the analogous decays in the $B$ sector, radiative
and leptonic modes such as $D^0 \to \rho \gamma$, $\D^+ \to \pi^+ l^+l^-$ and 
$D^0 \to \rho l^+l^-$
are of interest.  (At the time of writing the only `rare' decay of this
sort which has been observed is the channel $D^0 \to \phi \gamma$,
with a branching ratio of $2.5^{+0.7}_{-0.6}\times 10^{-5}$~\cite{DPHIGAM}.)
In general, however, although new physics contributions can be significantly larger
than the SM short-range expectations, it is almost certain that the
long-distance contributions are often completely dominant,  making 
the absolute branching ratios an unreliable indicator of 
beyond-the-SM effects.  Nevertheless, many channels hold their 
interest as the kinematical distributions, such as the dilepton invariant mass,
or the forward-backward asymmetry, retain their ability to discriminate
between the SM and new physics~\cite{RARECHARM1}.  
An example is provided by the dilepton invariant mass distribution
in $D^+ \to \pi^+ l^+l^-$ which has recently been proposed~\cite{LEPTOQ} as way in which to 
test whether leptoquarks can be invoked to explained the tension,
discussed in Sec.~\ref{sec:lattice}, between the lattice and experimental
determinations of $f_{D_s}$.

\section{Conclusions}

After several years of unjust neglect, charm physics is once more 
recognised as a discipline with a great deal to contribute to the 
future of HEP.
This sea-change has arisen through three unrelated 
reasons:  recent, unexpected, observations
in spectroscopy;   the realisation that measurements of $D$ decay
properties are essential technical inputs in the area of 
precision CKM physics; and the appreciation, given impetus by the
discovery of $D^0-\bar{D^0}$ mixing, that charm has its own
unique discovery potential.

\bigskip
I am grateful to Matt Shepherd for giving many useful suggestions during
the preparation of this talk,  and the organisers of PIC 2008 in Perugia for 
a stimulating conference in a beautiful location.

\end{document}